\documentstyle[epsfig]{mn2e}

\def\ang{\,{\rm\AA}}
\def\bj{b_{\scriptsize\rm J}}
\def\rf{r_{\scriptsize\rm F}}

\begin{document}

\title[E+A galaxies in the 2dFGRS]{The 2dF Galaxy Redshift Survey: the
local E+A galaxy population}

\author[Chris Blake et al.]{\parbox[t]{\textwidth}{Chris Blake$^1$,
Michael B.\ Pracy$^1$, Warrick J.\ Couch$^1$, Kenji Bekki$^1$, Ian
Lewis$^2$, Karl Glazebrook$^9$, Ivan K.\ Baldry$^9$, Carlton M.\
Baugh$^4$, Joss Bland-Hawthorn$^3$, Terry Bridges$^3$, Russell
Cannon$^3$, Shaun Cole$^4$, Matthew Colless$^5$, Chris Collins$^6$,
Gavin Dalton$^{2,15}$, Roberto De Propris$^5$, Simon P.\ Driver$^5$,
George Efstathiou$^7$, Richard S.\ Ellis$^8$, Carlos S.\ Frenk$^4$,
Carole Jackson$^{16}$, Ofer Lahav$^7$, Stuart Lumsden$^{10}$, Steve
Maddox$^{11}$, Darren Madgwick$^{13}$, Peder Norberg$^{14}$, John A.\
Peacock$^{12}$, Bruce A.\ Peterson$^5$, Will Sutherland$^{12}$, Keith
Taylor$^8$} \vspace*{12pt} \\ $^1$School of Physics, University of New
South Wales, Sydney, NSW 2052, Australia \\ $^2$Astrophysics,
University of Oxford, Keble Road, Oxford, OX1 3RH, UK \\
$^3$Anglo-Australian Observatory, P.O.\ Box 296, Epping, NSW 2111,
Australia\\ $^4$Department of Physics, University of Durham, South
Road, Durham DH1 3LE, UK \\ $^5$Research School of Astronomy \&
Astrophysics, The Australian National University, Weston Creek, ACT
2611, Australia \\ $^6$Astrophysics Research Institute, Liverpool John
Moores University, Twelve Quays House, Birkenhead, L14 1LD, UK \\
$^7$Institute of Astronomy, University of Cambridge, Madingley Road,
Cambridge CB3 0HA, UK \\ $^8$Department of Astronomy, California
Institute of Technology, Pasadena, CA 91025, USA \\ $^9$Department of
Physics \& Astronomy, Johns Hopkins University, Baltimore, MD
21118-2686, USA \\ $^{10}$Department of Physics, University of Leeds,
Woodhouse Lane, Leeds, LS2 9JT, UK \\ $^{11}$School of Physics \&
Astronomy, University of Nottingham, Nottingham NG7 2RD, UK \\
$^{12}$Institute for Astronomy, University of Edinburgh, Royal
Observatory, Blackford Hill, Edinburgh EH9 3HJ, UK \\ $^{13}$Lawrence
Berkeley National Laboratory, 1 Cyclotron Road, Berkeley, CA 94720,
USA \\ $^{14}$ETHZ Institut fur Astronomie, HPF G3.1, ETH Honggerberg,
CH-8093 Zurich, Switzerland \\ $^{15}$Rutherford Appleton Laboratory,
Chilton, Didcot, OX11 0QX, UK \\ $^{16}$CSIRO Australia Telescope
National Facility, PO Box 76, Epping, NSW 1710, Australia}

\maketitle

\begin{abstract}
We select a sample of low-redshift ($z \sim 0.1$) E+A galaxies from
the 2dF Galaxy Redshift Survey (2dFGRS).  The spectra of these objects
are defined by strong hydrogen Balmer absorption lines (H$\delta$,
H$\gamma$, H$\beta$) combined with a lack of [OII] $3727 \ang$
emission, together implying a recently-truncated burst of star
formation.  The E+A spectrum is thus a signpost to galaxies in the
process of evolution.

We quantify the local environments, clustering properties and
luminosity function of the E+A galaxies.  We find that the
environments are consistent with the ensemble of 2dFGRS galaxies:
low-redshift E+A systems are located predominantly in the field,
existing as isolated objects or in poor groups.  However, the
luminosity distribution of galaxies selected using three Balmer
absorption lines H$\delta\gamma\beta$ appears more typical of
ellipticals.  Indeed, morphologically these galaxies are
preferentially spheroidal (E/S0) systems.  In a small but significant
number we find evidence for recent major mergers, such as tidal tails.
We infer that major mergers are one important formation mechanism for
E+A galaxies, as suggested by previous studies.  At low redshift the
merger probability is high in the field and low in clusters, thus
these recently-formed spheroidal systems do not follow the usual
morphology-density relation for ellipticals.

Regarding the selection of E+A galaxies: we find that basing the
Balmer-line criterion solely on H$\delta$ absorption leads to a
significant sub-population of disk systems with detectable H$\alpha$
emission.  In these objects the [OII] emission is presumably either
obscured by dust or present with a low signal-to-noise ratio, whilst
the (H$\gamma$, H$\beta$) absorption features are subject to
emission-filling.
\end{abstract}

\section{Introduction}

E+A galaxies possess a characteristic spectral signature defined by
strong hydrogen Balmer absorption lines (H$\delta$, H$\gamma$,
H$\beta$) combined with a lack of optical emission lines such as [OII]
$3727 \ang$ (Dressler \& Gunn 1983).  The Balmer absorption lines are
imprinted in the galaxy spectrum by A stars, indicating the presence
of a young ($< 1$ Gyr old) stellar population.  However, the absence
of [OII] $3727 \ang$ emission suggests that star formation is no
longer ongoing.  The inference is that these galaxies have previously
undergone a burst of star formation, which has recently been truncated
rather suddenly (Dressler \& Gunn 1983; Couch \& Sharples 1987).  For
this reason, these systems are also known as `post-starburst
galaxies'.

The physical mechanisms governing the triggering and cessation of this
starburst are not yet fully understood, but undoubtedly reflect the
interaction of the galaxy with its environment.  In general, this can
occur through interaction with other galaxies (mergers or tidal
gravitational effects) or, for those galaxies residing in clusters,
through effects specific to that environment; involving either the hot
intracluster gas (ram-pressure effects) or the cluster gravitational
potential.  Theoretical modelling of all these scenarios is not yet
complete, although it is known that a major galaxy merger can produce
the characteristic E+A spectral signature (Bekki, Shioya \& Couch
2001).  In practice, more than one of the above mechanisms is probably
responsible for the overall population of E+A galaxies, with different
mechanisms operating in different environments.  As such, these
galaxies are an interesting probe of environmental influences on
galaxy evolution.

The population of E+A galaxies exhibits dramatic evolution with
redshift.  These galaxies are commonplace in intermediate-redshift
clusters, where they were first identified and studied (Dressler \&
Gunn 1983): `E' and `A' representing respectively the assumed
morphology and the dominant stellar lines of the spectrum, although
high-resolution Hubble Space Telescope (HST) imaging of such clusters
later revealed that these galaxies were predominantly early-type disk
systems (Couch et al.\ 1998; Dressler et al.\ 1999).  Whilst E+A
galaxies are prevalent in intermediate-redshift clusters (e.g.\ Tran
et al.\ 2003), they only constitute about 1 per cent of the members of
nearby clusters (Fabricant, McClintock \& Bautz 1991).  As a fraction
of the overall zero-redshift galaxy population, E+A objects comprise
significantly less than 1 per cent (Zabludoff et al.\ 1996).  This
rarity has rendered environmentally-unbiased studies of low-redshift
E+A galaxies in statistically significant numbers difficult until the
recent advent of large-scale galaxy surveys.

The first such study was that of Zabludoff et al.\ (1996) who
identified 21 low-redshift ($0.05 < z < 0.13$) E+A systems from 11113
galaxies in the Las Campanas Redshift Survey.  Interestingly, 75 per
cent of these objects were located in the field, well outside the rich
clusters in which E+A galaxies were originally studied, implying that
a cluster-specific mechanism is not essential for E+A galaxy
formation.  Furthermore, the ground-based morphologies of these
galaxies showed evidence for tidal features in 5 out of the 21 cases,
implicating galaxy-galaxy mergers or interactions as a formation
mechanism.  HST imaging (Yang et al.\ 2004) later strengthed these
hints by revealing the detailed morphological picture of a gas-rich
merger.  The Zabludoff et al.\ sample was also investigated via
spatially-resolved spectroscopy (Norton et al.\ 2001), probing the
kinematics of the component stellar populations, and providing further
evidence that E+A galaxies in the field represent a transitional phase
between disk-dominated, rotationally-supported systems and
spheroid-dominated, pressure-supported galaxies.

In this paper we extend the study of Zabludoff et al.\ (1996) by
selecting a low-redshift E+A sample from the 221000 galaxy spectra
which form part of the final data release of the 2dF Galaxy Redshift
Survey (2dFGRS; Colless et al.\ 2001, 2003).  We use two different
selection techniques, the first based on three Balmer absorption lines
(H$\delta$, H$\gamma$, H$\beta$) and the second utilizing solely the
H$\delta$ line.  These two methods yield samples of 56 and 243
galaxies respectively.  The significantly increased size of these
catalogues (with respect to Zabludoff et al.\@) allows us to measure
statistical properties of the E+A galaxy population, such as the
luminosity function and the clustering properties, and provides a
wider database for exploring morphologies.  We note that similar
low-redshift populations of E+A galaxies have been selected from the
Sloan Digital Sky Survey (SDSS): Goto et al.\ (2002) presented a
catalogue of H$\delta$-selected SDSS galaxies and Quintero et al.\
(2004) fitted a linear sum of A-star and K-star spectra to SDSS
galaxies, analyzing the photometric properties of the sub-sample with
an excess ratio of A-star to K-star components.  We compare our
results to this work where possible.

The plan of this paper is as follows: in Section \ref{secsamp} we
describe the selection of our 2dFGRS E+A galaxy samples, compare the
selected populations to theoretical tracks in the [colour,
EW(H$\delta$)]--plane, and use H$\alpha$ emission to test for any
ongoing but obscured star formation.  In Section \ref{secmorph} we use
Supercosmos Sky Survey images to investigate the E+A galaxy
morphologies.  In Sections \ref{secenv}, \ref{secclus} and \ref{seclf}
we measure various statistical properties of the E+A galaxy
population: the local environments, clustering properties and
luminosity function.  Throughout this paper, we convert redshifts to
physical distances using cosmological parameters $\Omega_m = 0.3$,
$\Omega_\Lambda = 0.7$, $h = H_0/(100 \; {\rm km} \; {\rm s}^{-1} \;
{\rm Mpc}^{-1}) = 0.7$.

\section{Sample selection}
\label{secsamp}

\subsection{The 2dF Galaxy Redshift Survey}

We selected E+A galaxies from the final data release of the 2-degree
Field Galaxy Redshift Survey (2dFGRS), a major spectroscopic survey of
about 221000 galaxies undertaken using the 2dF facility at the
Anglo-Australian Observatory.  The 2dFGRS covers an area of
approximately 1500 deg$^2$ in three regions: an North Galactic Pole
(NGP) strip, a South Galactic Pole (SGP) strip and a series of random
fields scattered around the SGP strip.  The 2dFGRS input catalogue was
selected in the photographic $\bj$ band from the Automatic Plate
Measuring facility (APM) galaxy survey (and its subsequent
extensions), with a nominal extinction-corrected magnitude limit $\bj
= 19.45$.  The survey spectra were obtained through 2-arcsecond fibres
and cover the wavelength range $3600-8000 \ang$ at a resolution of $9
\ang$.  This wide wavelength range is made possible by the use of an
atmospheric dispersion compensator (ADC) within the 2dF instrument.
Each spectrum is visually assigned a redshift quality flag $Q$ which
ranges from $Q=1$ (unreliable) to $Q=5$ (highest quality).  The
spectra are not flux calibrated and thus consist of a sequence of
`counts' as a function of wavelength.  The 2dFGRS is described in
detail in Colless et al.\ (2001, 2003) and the database may be
accessed online at {\tt http://msowww.anu.edu.au/2dFGRS/}.

\subsection{The 2dFGRS spectral line catalogue}
\label{seclinefit}

Our sample selection is based on the 2dFGRS spectral line catalogue
prepared by Ian Lewis.  In this Section we briefly summarize the
generation of the spectral line catalogue; for full details see Lewis
et al.\ (2002).  After removal of the continuum by subtracting the
median over windows of width $133 \ang$, Lewis et al.\ fitted up to
twenty absorption or emission lines, corresponding to standard galaxy
spectral features.  The line profiles were assumed to be Gaussian,
parameterized by an amplitude and width.  The wavelength spacings of
the line centres were fixed at their known laboratory values, with a
variable overall offset to accommodate redshifting.  The quality of
each line fit was classified by a flag determined by the rms
residuals, ranging from 0 (bad fit) to 5 (good fit), and a
signal-to-noise parameter was computed for each line (relative to the
continuum).  Where possible, equivalent widths were then deduced using
the fitted line flux and the value of the continuum flux in the local
$133 \ang$ window.  This does not require absolute flux calibration of
the spectra, although we must assume that there is no significant
additive error in the continuum due to effects such as scattered
light.  We corrected all equivalent widths for cosmological effects by
dividing by a factor $(1+z)$, where $z$ is the galaxy redshift.

\subsection{E+A galaxy selection criteria}

Galaxies exhibit a continuum of properties and thus the choice of
selection criteria for a specific sub-class is somewhat arbitrary.
Zabludoff et al.\ (1996) performed the first environmentally-unbiased
selection of E+A galaxies, obtained from the Las Campanas Redshift
Survey (LCRS).  Their sample was defined by requiring the equivalent
width of [OII] $3727 \ang$ emission to be less than $2.5 \ang$, and
the average of the equivalent widths of the Balmer lines H$\delta$,
H$\gamma$ and H$\beta$ to exceed $5.5 \ang$ in absorption.  These
criteria are strict, selecting an extreme class of objects
corresponding to 0.2 per cent (21/11113) of LCRS galaxies.  Zabludoff
et al.\ only considered spectra possessing a signal-to-noise ratio
exceeding $8.0$ in the continua about the H$\delta$, H$\gamma$ and
H$\beta$ lines.

We based our E+A galaxy sample selection criteria on those of
Zabludoff et al.\@, but with several adjustments.  Firstly, the 2dFGRS
line fits catalogue (Section \ref{seclinefit}) parameterizes the
quality of the fit to each spectral line using a different method to
that of Zabludoff et al.\@; thus we cannot employ an identical
selection criterion as regards signal-to-noise ratio.  In our analysis
we considered an equivalent width measurement to be reliable if the
signal-to-noise parameter of the line exceeded $1.0$ and the flag
parameter of the line was equal to 4 or 5 (these flags are classified
as a `good' fit).  Note that a `signal-to-noise parameter' equal to
$1.0$ is in fact a high-quality detection: this parameter is defined
as the mean signal-to-noise ratio of the resolution elements of the
line, averaged over three line-widths.  The choice of a minimum value
of $1.0$ is fairly arbitrary, but serves to select an extreme class of
galaxies as required.

\begin{figure}
\center
\epsfig{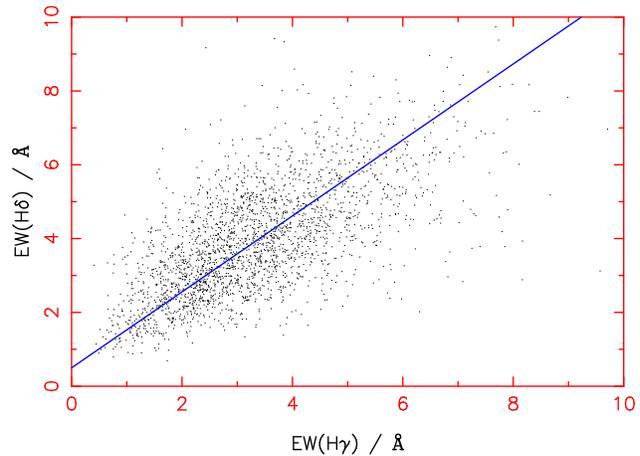}
\caption{Distribution of equivalent widths in absorption of H$\gamma$
and H$\delta$ for the 2dFGRS spectral line catalogue.  Galaxies are
only plotted if they possess redshift $z > 0.002$, high-quality 2dFGRS
spectra (defined by redshift quality flag $Q \ge 3$ and ADC flag $=
1$) and `good' measurements of each of these equivalent widths
(defined by a signal-to-noise parameter $> 1.0$ and a flag parameter
$\ge 4$).  The straight line is a minimum chi-squared linear fit to
the data.  This is generated by assuming an error of $1 \ang$ in both
equivalent widths and using a standard fitting routine.  We restricted
the fit to equivalent widths between $0 \ang$ and $10 \ang$ in
absorption.}
\label{fighghd}
\end{figure}

Furthermore, rather than using the average of Balmer equivalent
widths, we introduced a weighting scheme.  The relative equivalent
widths of the Balmer absorption lines (H$\delta\gamma\beta$) in a
given galaxy spectrum are determined partly by fundamental atomic
physics and are hence expected to display some level of correlation.
We fitted empirical linear relations to the distributions of
(H$\gamma$,H$\delta$) and (H$\beta$,H$\delta$) equivalent widths for
objects with reliable measurements of these lines in absorption (see
Fig.\ \ref{fighghd} for more details of our fitting procedure).  We
thereby derived empirical best-fitting correlations:
\begin{equation}
{\rm EW(H}\delta) = 0.50 + 1.03 \times {\rm EW(H}\gamma)
\end{equation}
\begin{equation}
{\rm EW(H}\delta) = -3.53 + 1.67 \times {\rm EW(H}\beta)
\end{equation}
where all equivalent widths are measured in $\ang$ in absorption.  We
used these linear fitting formulae to convert H$\beta$ and H$\gamma$
equivalent widths to `effective' H$\delta$ values.  For each spectrum
we averaged these values for the H$\delta\gamma\beta$ absorption
lines.  Fig.\ \ref{fighghd} displays the scatter plot for the case of
(H$\gamma$,H$\delta$).

We selected catalogues of E+A galaxies using two different methods;
this allowed us to ascertain the effects of the selection criteria on
the properties of the sample.  The first selection method was based on
the average of the weighted Balmer `effective H$\delta$' equivalent
widths described above, the value of which was required to exceed $5.5
\ang$ in absorption.  The second selection was based solely on the
H$\delta$ equivalent width, which was required to exceed $5.5 \ang$.
We emphasize again that as there is no sign of bimodality in the
sample, the choice of the value $5.5 \ang$ is fairly arbitrary.  All
E+A objects were required to have no detection of [OII] $3727 \ang$
emission.  Non-detection was defined either by an [OII] flag parameter
equal to zero, or by a flag parameter $\ge 4$ and an [OII] equivalent
width less than $2.5 \ang$ in emission.

In more detail: the catalogue of 2dFGRS spectral line fits compiled by
Lewis et al.\ (2002) contains 264765 spectra, including repeat
observations of the same source.  We only used high-quality 2dFGRS
galaxy spectra defined by:
\begin{itemize}
\item 2dFGRS redshift quality flag $Q \ge 3$
\item ADC flag $= 1$ (spectra observed earlier than August 1999 are
afflicted by problems with the ADC, resulting in severe depletion of
counts at the blue end of the spectrum, and are assigned ADC flag $=
0$)
\item Redshift $z > 0.002$ (in order to exclude stars)
\end{itemize}
These cuts left a total of 161437 high-quality galaxy spectra.  We
next required that there was little or no detectable [OII] $3727 \ang$
line emission, defined by either
\begin{itemize}
\item Flag parameter of [OII] $3727 \ang = 0$
\end{itemize}
or
\begin{itemize}
\item Flag parameter of [OII] $3727 \ang \ge 4$
\item Equivalent width of [OII] $3727 \ang$ emission $< 2.5 \ang$
\end{itemize}
This requirement was satisfied by 66422 galaxies.  For the first of
our two E+A galaxy catalogues, we made the selection for Balmer
absorption based on the H$\delta\gamma\beta$ lines:
\begin{itemize}
\item Signal-to-noise parameter of each of H$\delta\gamma\beta > 1.0$
\item Flag parameter of each of H$\delta\gamma\beta \ge 4$
\item The average `effective H$\delta$' equivalent width (where
H$\gamma\beta$ equivalent widths are weighted by the fitted conversion
factors) $> 5.5 \ang$ in absorption
\end{itemize}
This produced a final sample of 56 galaxies (0.03 per cent of
high-quality galaxy spectra).  We refer to this sample as the
`average-Balmer E+A catalogue' (Table \ref{tabavebalm}).

\begin{figure*}
\center
\epsfig{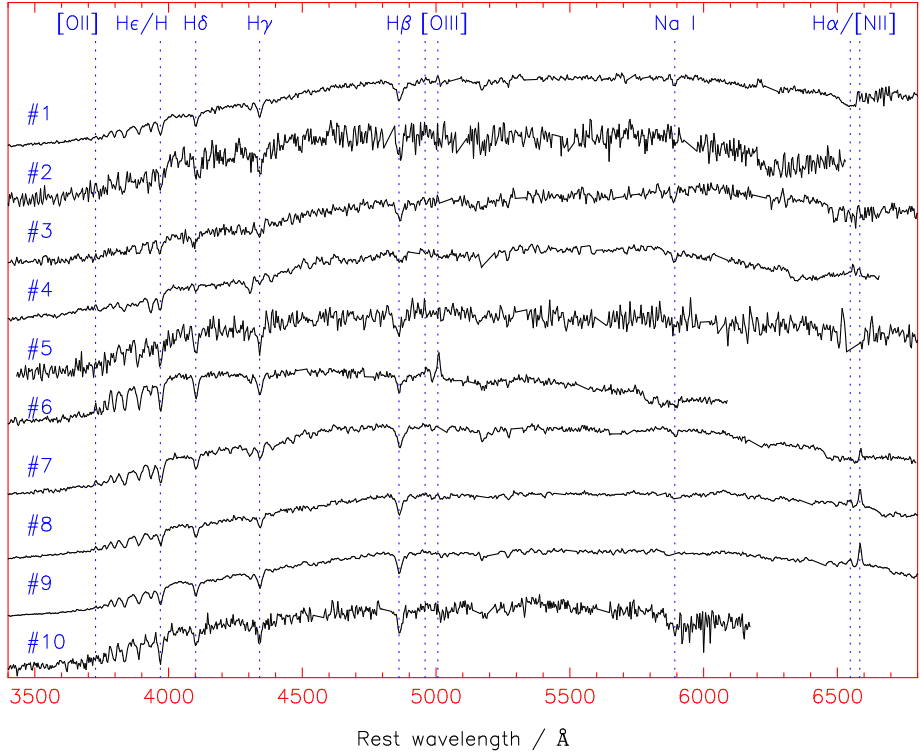}
\caption{2dFGRS spectra of the first ten objects in the average-Balmer
E+A catalogue (Table \ref{tabavebalm}), with wavelengths corrected to
the galaxy rest frame.  Positions of important spectral features are
indicated by the vertical dotted lines.  The wavelength ranges of
prominent night-sky emission or absorption features have been excised
and interpolated over.}
\label{figepaabspec}
\end{figure*}

Fig.\ \ref{figepaabspec} plots the first ten 2dFGRS spectra from this
catalogue, with wavelengths corrected to the galaxy rest frame (there
is no bias induced by plotting the first ten; this is a representative
sub-sample).  Important spectral features are indicated.  Weak
emission lines are occasionally evident ([OIII] $5007 \ang$ in
spectrum $\#6$ and [NII] $6584 \ang$ in $\#8$ and $\#9$) that are
potentially due to weak star formation (see Section \ref{sechalpha})
or to AGN activity.

\begin{figure*}
\center
\epsfig{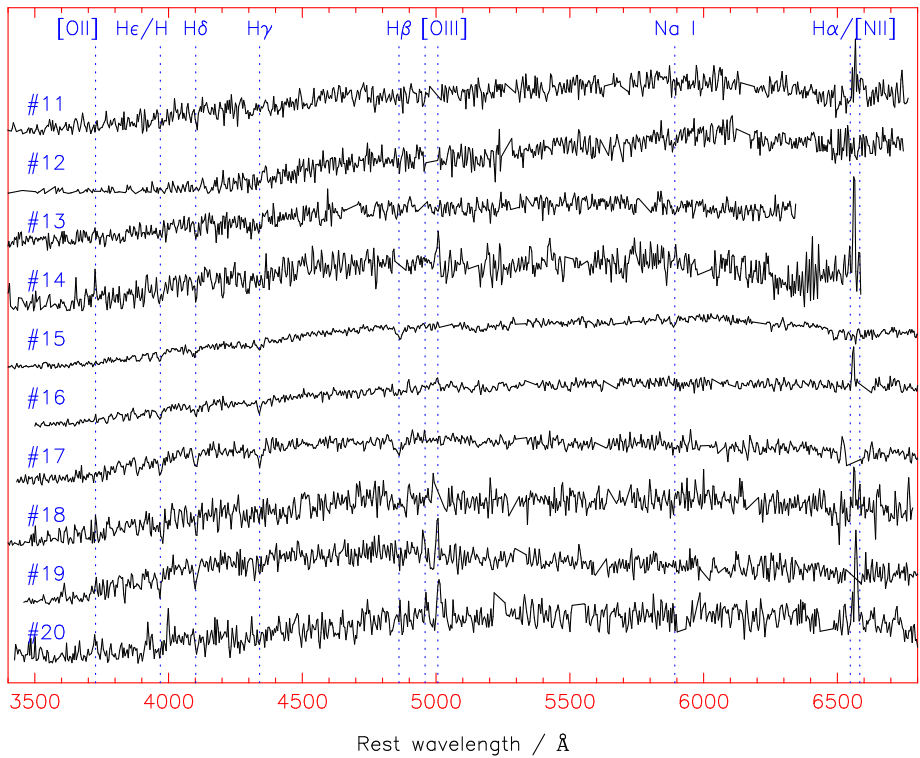}
\caption{2dFGRS spectra of the first ten objects in the H$\delta$ E+A
catalogue, displayed in the same manner as Fig.\ \ref{figepaabspec}.}
\label{figepahdspec}
\end{figure*}

We note that the H$\beta$ line is fitted with both an emission and an
absorption Gaussian component in the spectral line catalogue.  We only
considered the absorption component; in almost all relevant cases the
emission component was negligible.

For the second of our two E+A galaxy catalogues, we made the Balmer
absorption line selection based purely on the equivalent width of
H$\delta$:
\begin{itemize}
\item Signal-to-noise parameter of H$\delta > 1.0$
\item Flag parameter of H$\delta \ge 4$
\item H$\delta$ equivalent width $> 5.5 \ang$ in absorption
\end{itemize}
This resulted in a sample of 243 galaxies (0.15 per cent of
high-quality spectra).  We refer to this sample as the `H$\delta$ E+A
catalogue'.  We note that this sample contains 36 of the 56
average-Balmer catalogue members.

Fig.\ \ref{figepahdspec} plots the first ten 2dFGRS spectra from this
catalogue.  We note that selecting objects by requiring a significant
detection of solely H$\delta$, rather than three Balmer absorption
lines, results in more examples of spectra with low median
signal-to-noise ratios, in which the measured H$\delta$ equivalent
width may be an over-estimate due to noise.  Interestingly, a
significant fraction of spectra ($\#11$, $\#14$, $\#16$, $\#20$)
contain H$\alpha$ emission indicative of ongoing star formation, even
though no measurement of [OII] $3727 \ang$ was possible.  We discuss
this issue further in Section \ref{sechalpha}.

Unsurprisingly, in the presence of noise, a significantly larger
subset of objects results if we stipulate a good measurement of only
one Balmer absorption line rather than of three.  We consider that the
average-Balmer E+A catalogue provides the highest-fidelity selection
of `true E+A galaxies'.  This statement is justified further below.
We note that both selection techniques used here are stricter than
that of Zabludoff et al.\ (1996) in terms of the fraction of objects
chosen from the parent catalogue.  This is due to the different
requirements on signal-to-noise ratios in the two analyses.

\subsection{Colours}
\label{seccol}

\begin{figure}
\center
\epsfig{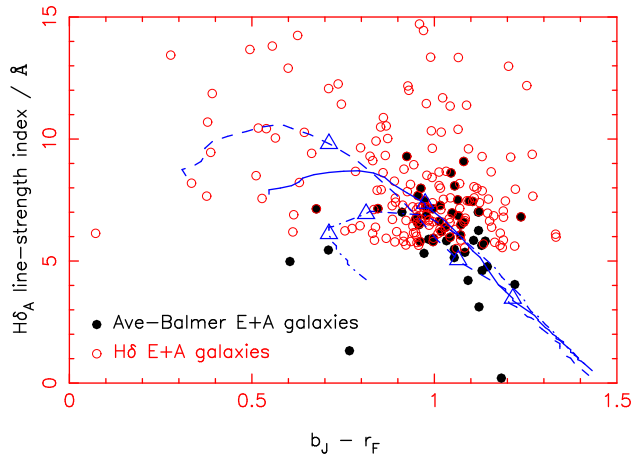}
\caption{The H$\delta_{\rm \scriptsize A}$ line-strength index
(Worthey \& Ottaviani 1997) and photographic $\bj - \rf$ colours of
objects in the two E+A galaxy catalogues.  The overplotted lines are
evolutionary tracks calculated using the stellar population synthesis
models of Bruzual \& Charlot (2003), as described in the text.  The
two large plotted triangles on each track indicate time points 0.5 Gyr
and 1 Gyr after the starburst.  The errors in the line-strength index
measurements (omitted for clarity) are typically $0.5 \rightarrow 1
\ang$ for the average-Balmer E+A catalogue, but can range up to $4
\ang$ for the H$\delta$ E+A catalogue (particularly for measured line
strengths exceeding $10 \ang$).}
\label{figepacol}
\end{figure}

Fig.\ \ref{figepacol} plots the H$\delta$ equivalent widths of objects
in the two E+A galaxy catalogues against their photographic $\bj -
\rf$ colour.  The equivalent widths were re-measured for these samples
using the H$\delta_{\rm \scriptsize A}$ line-strength index defined by
Worthey \& Ottaviani (1997) in place of the Gaussian-fitting result,
to facilitate a consistent comparison with theoretical models.  The
$\bj$ and $\rf$ magnitudes of the 2dFGRS galaxies are listed in the
publicly-accessible 2dFGRS online database; we used the Supercosmos
Sky Survey magnitudes.  In generating Fig.\ \ref{figepacol} we
K-corrected these magnitudes to their rest-frame values using the
formulae listed in Wild et al.\ (2004), which produce K-corrections
appropriate to each 2dFGRS galaxy's redshift $z$ and colour, and which
are accurate to $0.01$ mag in almost all cases (ignoring galaxy
evolution):
\begin{equation}
K_b = [-1.63+4.53 x] \, y + [-4.03-2.01 x] \, y^2 - \frac{z}{1+(10z)^4}
\end{equation}
\begin{equation}
K_r = [-0.08+1.45 x] \, y + [-2.88-0.48 x] \, y^2
\end{equation}
where $x = \bj - \rf$ and $y = z/(1+z)$.

It is interesting to observe in Fig.\ \ref{figepacol} that despite
their starburst origin, not all our E+A galaxies are blue, but rather
cover quite a broad range in colour: for both E+A catalogues, the mean
value of $\bj-\rf$ is within $0.05$\,mag of that of the entire 2dFGRS.
The reddest E+A galaxies are only $\sim 0.3$\,mag bluer than a passive
E/S0 galaxy [$(\bj-\rf)_{z=0} \simeq 1.6$].  Indeed their distribution
in the [colour, EW(H$\delta$)]--plane is similar to that of the E+A
populations observed in distant clusters (e.g.\ Couch \& Sharples
1987; Dressler et al.\ 1998), apart from our sample not extending all
the way to the E/S0 fiducial colour.

Extensive modelling of this distribution observed for E+A galaxies in
distant clusters has shown that it is best reproduced by starburst
models, where the galaxies are seen at different times after a
substantial but abruptly truncated episode of star formation (Barger
et al.\ 1996; Poggianti et al.\ 1999). Models of this type have been
run for our present sample, and these are represented by the various
tracks plotted in Fig.\ \ref{figepacol}.  Here we have used the
stellar population synthesis models of Bruzual \& Charlot (2003).
Their code outputs at a series of time-steps the $B - R$ colour and
H$\delta_{\rm \scriptsize A}$ line-strength index, computed directly
from high-resolution model spectra; we transformed the outputted $B -
R$ colour into a photographic $\bj - \rf$ colour using the equation
(Couch 1981):
\begin{equation}
\bj - \rf = -0.017 + 1.059 \, (B-R) - 0.027 \, (B-R)^2
\end{equation}

The results are overplotted in Fig.\ \ref{figepacol} for various cases
of interest.  In all models we assumed an exponentially-decaying
`background' rate of star formation with e-folding time $\tau = 3$ Gyr
(our results do not depend significantly on the value of $\tau$).  At
time $t = 10$ Gyr we superimposed a $\delta$-function starburst
creating 10 per cent of the total stellar mass of the model galaxy.
The solid line in Fig.\ \ref{figepacol} tracks the subsequent
evolution of this galaxy from $t = 10.1$ Gyr to $t = 13$ Gyr.  The
other two lines plot variations in the model: the dashed line
corresponds to the (extreme) assumption that the $\delta$-function
starburst forms 50 per cent of the stellar mass of the galaxy; the
dot-dashed line replaces the $\delta$-function starburst with a `flat'
starburst of duration 1 Gyr (creating 10 per cent of the total stellar
mass).  Invoking a reasonable model of dust extinction within Bruzual
\& Charlot's code (Charlot \& Fall 2000) reddens the $\bj - \rf$
colour of these tracks by $\approx 0.1$ mag, using the default dust
model parameters.

It is of note that a small number of observed galaxies in Fig.\
\ref{figepacol} (viz.\@, the reddest E+A galaxies with the strongest
H$\delta$ absorption) cannot be accounted for by the model tracks
through reasonable variations in the starburst pattern or in the dust
model.  For example, the reddest average-Balmer E+A galaxy possesses
colour $\bj - \rf = 1.24$ and line-strength index $6.8 \pm 0.2 \ang$
(although those line-strength measurements exceeding $10 \ang$ for
some H$\delta$-selected galaxies have associated errors ranging up to
$4 \ang$).  This difficulty that the starburst models have in
reproducing the colours and H$\delta$ equivalent widths of some `red
H$\delta$-strong' galaxies was first recognised by Couch \& Sharples
(1987) in their study of distant clusters, and to this day the
explanation has remained an unsolved puzzle.  The only viable
possibilities in this context would seem to be very heavy dust
extinction and/or an unusual stellar initial mass function (Shioya,
Bekki \& Couch 2004).

\subsection{Obscured star formation?}
\label{sechalpha}

In some cases, ongoing star formation in E+A galaxies may be hidden by
dust obscuration.  Smail et al.\ (1999) discovered examples of distant
cluster galaxies that were classified as E+A objects on the basis of
their optical spectra, yet possessed radio fluxes indicative of
current massive star formation.  However, only 2 out of 15 E+A
galaxies in the Zabludoff et al.\ (1996) sample could be detected in
radio continuum measurements (Miller \& Owen 2001; the radio continuum
luminosity is a tracer of ongoing star formation that is unbiased by
dust).  Furthermore, just 1 out of 5 of these objects yielded a
detection of neutral hydrogen gas via 21 cm emission (Chang et al.\
2001; the non-detected systems have upper limits of neutral hydrogen
content of $1 \rightarrow 2 \times 10^9 \, h^{-2} \, M_\odot$).  The
presence of neutral hydrogen would indicate the existence of a gas
reservoir that may fuel ongoing or future star formation (and the
large-scale spatial distribution and kinematics of this gas can encode
information about any merger event).  This evidence suggests that
low-redshift E+A galaxies are predominantly not undergoing
dust-enshrouded star formation.

In order to test this assertion further, for each E+A galaxy spectrum
in our 2dFGRS samples we inspected the equivalent width of H$\alpha$
emission, which is less sensitive to extinction by dust than [OII]
$3727 \ang$ emission.  For the purpose of this Section, we assume that
any H$\alpha$ emission is caused by star formation activity, although
we note that the presence of an AGN is an alternative explanation and
more detailed consideration of spectral line ratios is required to
distinguish between these two cases (e.g.\ Kauffmann et al.\ 2003).
As a second caveat, we note that we are subject to aperture effects
owing to the $2''$ angular diameter of the 2dF spectrograph fibres: we
may not be measuring the correct global H$\alpha$ equivalent width of
the entire galaxy.

Although the 2dFGRS spectra are not flux-calibrated, we can still
deduce a star-formation rate from an H$\alpha$ equivalent width if we
use each galaxy's $\rf$ magnitude to estimate the continuum flux at
about $6000 \ang$.  In detail, we transformed the $\rf$ magnitude to
an AB-magnitude using the conversions stated in Cross et al.\ (2004,
equations A12 and A16), where the effective wavelength of the AB
magnitude is $\lambda_{\rm eff} = 5595 \ang$ (Blanton et al.\ 2003).
The expression for the continuum luminosity per unit wavelength is
then $L_{\rm cont} = (4.18 \times 10^{24}$ W $\ang^{-1}) \times
10^{-0.4 M}$ (using e.g.\ Blanton et al.\ 2003, equation 15) where $M$
is the absolute AB magnitude.  The H$\alpha$ luminosity follows as
$L_{{\rm H} \alpha} = L_{\rm cont} \times$ EW(H$\alpha$), and finally
we use a standard conversion to star formation rate, $SFR (M_\odot$
yr$^{-1}) = 8.2 \times 10^{-35} L_{{\rm H} \alpha} ({\rm W})$ (e.g.\
Sullivan et al.\ 2001, equation 1).

We considered a galaxy spectrum to have a reliable H$\alpha$
measurement if the line possessed a signal-to-noise parameter $> 1.0$
and a flag parameter $\ge 4$.  We note that H$\alpha$ measurements are
only possible for galaxies with redshifts $z < 0.15$, owing to the
poor sensitivity of the 2dF system beyond about $7500 \ang$ (see
Figs.\ \ref{figepaabspec} and \ref{figepahdspec}).

In the average-Balmer E+A catalogue, of the 37 galaxy spectra with $z
< 0.15$, only three yielded a reliable measurement of H$\alpha$
emission, with the highest deduced star-formation rate being $0.17 \,
M_\odot$ yr$^{-1}$.  In strong contrast, of the H$\delta$-selected
sample, 96 of the 166 spectra with $z < 0.15$ contained measurable
H$\alpha$ emission, although in only 5 cases did the star formation
rate exceed $1 \, M_\odot$ yr$^{-1}$ (with the highest rate being $1.9
\, M_\odot$ yr$^{-1}$).  We conclude that the average-Balmer method is
considerably more effective in selecting a sample with negligible
H$\alpha$ emission, that has truly ceased star formation (this result
is also evident by comparing Figs.\ \ref{figepaabspec} and
\ref{figepahdspec}).  E+A galaxies selected on the basis of H$\delta$
absorption alone show evidence for measurable ongoing star formation
in $\approx 60$ per cent of cases; the [OII] $3727 \ang$ emission
presumably being either obscured by dust or present with a low
signal-to-noise ratio.  The magnitude of the star-formation rates in
these objects is not large, being consistent with the pedestrian
levels observed in spiral disks.  The H$\gamma$ and H$\beta$
absorption features are presumably subject to emission-filling and the
object is consequently not selected in the average-Balmer catalogue.
The H$\delta$ E+A catalogue may thus contain an additional galaxy
population: dusty disk galaxies.

We also checked for inclusion of E+A galaxy catalogue members in the
list of 2dFGRS radio sources compiled by E.Sadler (priv. comm.) after
careful cross-matching of the 2dFGRS catalogue and the $1.4$ GHz NRAO
VLA Sky Survey (NVSS) (Sadler et al.\ 2002).  The NVSS is complete to
a radio flux-density limit of $S_{\rm 1.4 \, GHz} \approx 3$ mJy,
corresponding to a radio luminosity $L_{\rm 1.4 \, GHz} \approx 6
\times 10^{21}$ W Hz$^{-1}$ at the typical redshift of a 2dFGRS
galaxy, $z = 0.1$ (assuming a power-law radio spectrum $S_\nu \propto
\nu^{-0.8}$).  This level of radio continuum flux density is generated
by a star-formation rate of about $7 \, M_\odot$ yr$^{-1}$ (using the
conversion stated in Sullivan et al.\ 2001, equation 3), therefore the
H$\alpha$ emission is a probe of lower star formation rates in the
present study.  In the average-Balmer (or H$\delta$) E+A catalogue,
just 1 out of 56 (2 out of 243) objects was listed as an NVSS radio
detection.

\section{Morphologies}
\label{secmorph}

We inspected Supercosmos Sky Survey (SSS) images of objects in our E+A
galaxy catalogues.  The SSS has digitized sky survey plates taken with
the UK Schmidt telescope, using a pixel size of 0.67 arcseconds.
Three different photographic colours are available (approximating B, R
and I) and the data were accessed using the website {\tt
http://www-wfau.roe.ac.uk/sss/pixel.html}.  An introduction to the SSS
is presented by Hambly et al.\ (2001).

Table \ref{tabmorph} displays the results of our visual morphological
classification of our E+A galaxy samples, based on inspection of the
blue SSS images.  All the galaxies in the average-Balmer catalogue
were classified, and a comparable number (selected at random) from the
H$\delta$ sample were also classified.  Galaxies were assigned one of
the following broad `Hubble' types: E, E/S0, S0, Spiral (early),
Spiral (late), or Irr(egular); unless they were too distant to be
sufficiently-well resolved.  Any evidence for tidal interactions and
mergers (e.g.\ tidal tails, coalescing pairs, disturbed appearance)
was also noted.

As can be seen from Table \ref{tabmorph}, {\it there is a clear
difference in the typical morphologies between these catalogues}.
Average-Balmer E+A galaxies are predominantly early-type spherodial
E/S0 systems.  There is evidence of a disk component in some objects,
but this contribution is never dominant.  In contrast,
H$\delta$-selected E+A galaxies are predominantly late-type spiral
galaxies.  This result is consistent with our analysis of the star
formation rates in Section \ref{sechalpha}: the H$\delta$ E+A
catalogue contains an additional population of star-forming disk
systems that is not present in the average-Balmer catalogue.  In this
sense, the latter catalogue provides a higher-fidelity selection of
`true E+A galaxies'.

\begin{table*}
\center
\caption{Results of a visual morphological classification of galaxies
in the two E+A catalogues, based on Supercosmos Sky Survey blue
photographic images.  The fractions of galaxies classified as types
(E, E/S0, S0, early spiral, late spiral, irregular) are shown,
together with the number of cases in which strong (or possible)
evidence was found for a previous or ongoing merger or interaction.}
\label{tabmorph}
\begin{tabular}{ccccccccccc}
\hline Sample & Number & Number & E & E/S0 & S0 & Spiral & Spiral &
Irregular & Strong evidence & Possible evidence \\ & inspected &
classified & & & & (early) & (late) & & for interaction & for
interaction \\ \hline Average-Balmer & 56 & 40 & $15\%$ & $40\%$ &
$23\%$ & $20\%$ & $0\%$ & $2\%$ & 5 ($9\%$) & 7 ($13\%$) \\ H$\delta$
& 71 & 57 & $14\%$ & $7\%$ & $11\%$ & $17\%$ & $44\%$ & $7\%$ & 3
($4\%$) & 4 ($6\%$) \\ \hline
\end{tabular}
\end{table*}

\begin{figure*}
\center
\epsfig{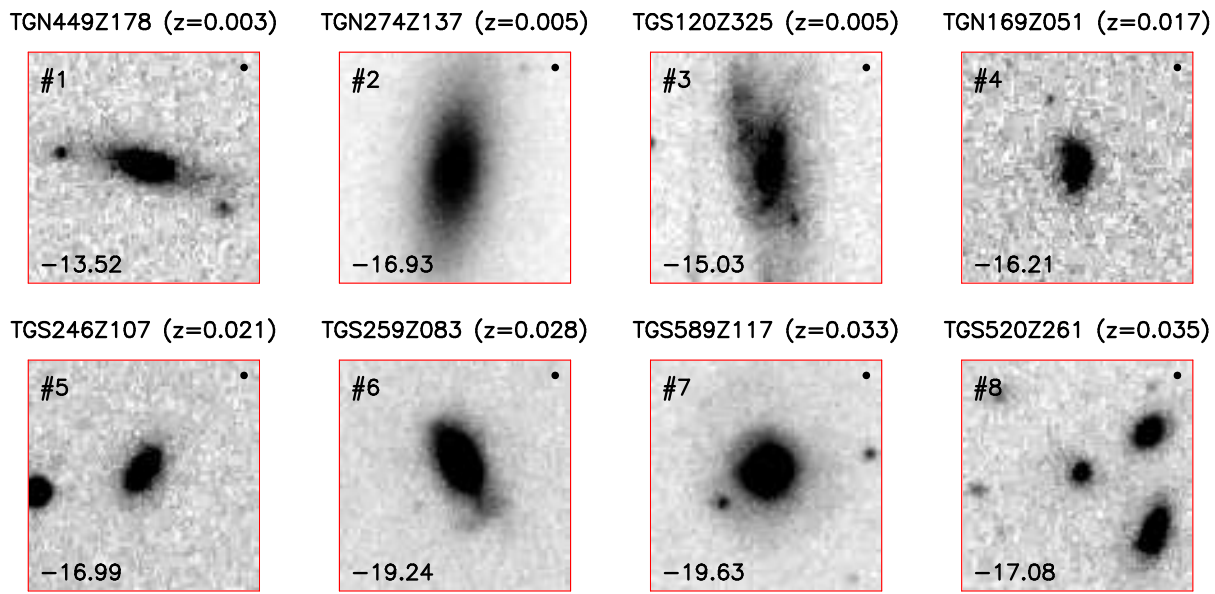}
\caption{Supercosmos Sky Survey images of the eight lowest-redshift
average-Balmer E+A galaxies (titled by 2dFGRS names).  The size of
each image is $1 \times 1$ arcmin, which corresponds to a co-moving
transverse width of $14$ kpc at redshift $z = 0.01$ for our assumed
cosmological parameters.  The pixel scale is 0.67 arcsec/pixel.  The
absolute magnitude of each galaxy is displayed in the bottom left-hand
corner of the plot (as the value of $M_b - 5 \, {\rm log_{10}} \, h$);
the small circle in the top right-hand corner indicates the diameter
of a 2dF spectrograph fibre (2 arcseconds, which corresponds to 0.4
kpc at $z = 0.01$).}
\label{figepaabimg}
\end{figure*}

Images of the eight lowest-redshift average-Balmer E+A galaxies are
displayed in Fig.\ \ref{figepaabimg}, using the blue SSS data to match
the selection colour of the 2dFGRS.  Note that there is no
morphological selection imposed for this Figure, only the original
spectral selection together with proximity.  The plots are labelled
with the absolute magnitude of each galaxy: these nearby objects are
in all cases fainter than $L^*$ ($M^*_b - 5 \, {\rm log_{10}} \, h =
-19.66$, Norberg et al.\ 2002).

\begin{figure*}
\center
\epsfig{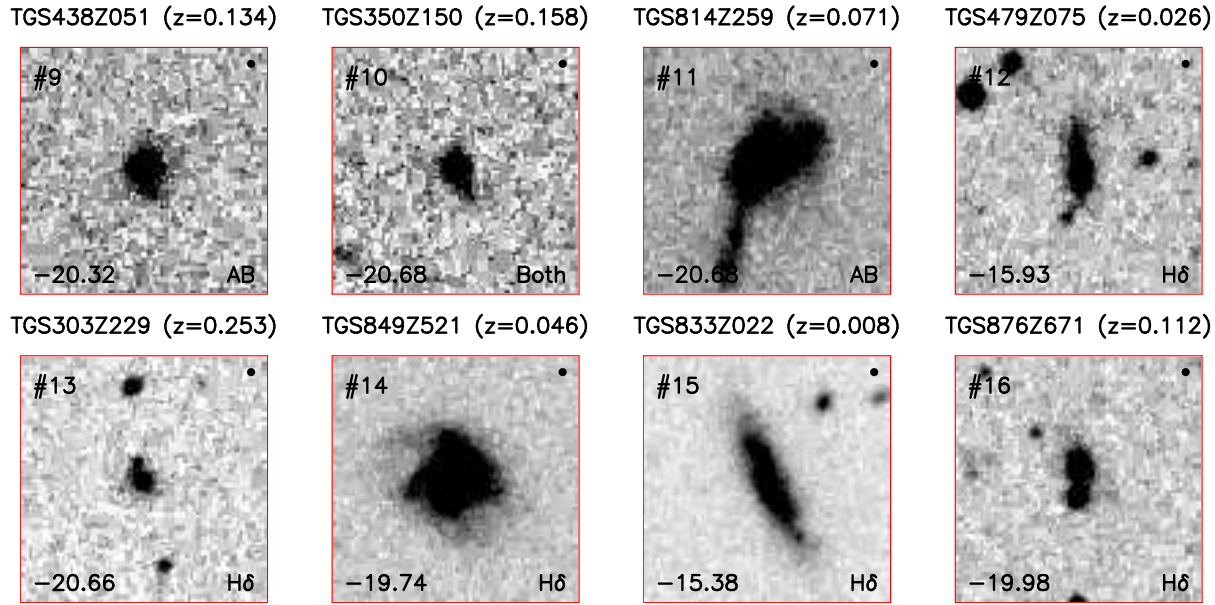}
\caption{Supercosmos Sky Survey images of eight possible merging or
interacting E+A galaxies, displayed in the same manner as Fig.\
\ref{figepaabimg}.  The bottom right-hand corner records the E+A
catalogue(s) to which each galaxy belongs: `AB' (average Balmer),
`H$\delta$' or `Both'.}
\label{figepamerge}
\end{figure*}

Our visual inspection of the SSS images also revealed that a small but
significant number of the E+A galaxies possess disturbed morphologies
or tidal features indicative of a late stage of a major merger, in
agreement with the findings of Zabludoff et al.\ (1996).  For example,
in Fig.\ \ref{figepaabimg} we classified images $\#3$ and $\#6$ as
providing `strong' evidence for a recent merger, and images $\#4$,
$\#7$ and $\#8$ as `possible' candidates (see the last two columns of
Table \ref{tabmorph}).  Fig.\ \ref{figepamerge} displays some other
examples of possible mergers or interactions drawn from both
catalogues; image $\#11$ features a bright tidal tail extending for a
physical distance of $\sim 50$ kpc.  Theoretical modelling has indeed
shown that major mergers can produce the characteristic E+A spectrum
(Bekki et al.\ 2001), and space-based high-resolution imaging of E+A
galaxies (Yang et al.\ 2004) has provided further support for this
scenario.  We note that merger events have a relatively short `duty
cycle' ($\sim 10^8$ yr) compared to the timescale of the E+A phase
($\sim 10^9$ yr): we expect morphological evidence for mergers to be
rare.

As already indicated, the H$\delta$-selected E+A galaxies display a
wider range of morphologies than the average-Balmer catalogue.  We
discover plausible merger remnants, but we also identify
low-luminosity disk systems which display no evidence of recent or
current interactions with other galaxies.  Major mergers cannot
produce such disk systems: one possible alternative formation
mechanism for such an E+A galaxy is a tidal interaction with a
companion that passed by $\sim 1$ Gyr ago.  Dwarf galaxies are
particularly susceptible to such tidal interactions, which can
transform a late-type spiral galaxy into a barred S0 galaxy that still
preserves the edge-on appearance of a disk system.  The dynamics of
the bar drives gas to the centre of the system, triggering a starburst
(e.g.\ Kennicutt 1998); the rapid consumption of this gas naturally
produces a characteristic E+A spectrum.

It is important to note that the angular diameter of each fibre of the
2dF spectrograph (2 arcseconds) is significantly smaller than the
angular size of most of these galaxies.  The E+A galaxy selection is
therefore subject to aperture effects: we can only assert that the
small portion of the galaxy sampled by the fibre possesses an E+A
spectrum.  This obviously does not preclude ongoing star formation in
other regions of the galaxy.  However, a starburst triggered by a
major galaxy merger or interaction will typically reside in the centre
of the system, where gas is driven by the dynamics: thus such a
starburst should co-exist with the highest optical surface brightness,
where the spectrograph fibre is usually positioned.

\section{Environments}
\label{secenv}

We quantified the local environments of the E+A galaxy samples using a
variety of techniques:

\subsection{Cross-correlation with the cluster distribution}

We first investigated whether our samples of low-redshift E+A galaxies
are located predominantly in the field, as found in the seminal study
of Zabludoff et al.\ (1996), or in the proximity of rich clusters.  A
catalogue of galaxy clusters within the 2dFGRS observed sky areas was
compiled by de Propris et al.\ (2002), sourced from the Abell, APM and
Edinburgh-Durham Cluster (EDCC) catalogues.  The 2dFGRS was utilized
by de Propris et al.\ to measure precise redshifts, velocity
dispersions and centroids for these clusters.

We simply measured the vector separation of each E+A galaxy and the
catalogued clusters.  Two components of separation were measured: the
transverse (i.e.\ projected) physical distance $D_t = (r \times
\Delta\theta)/(1+z)$, where $r$ is the radial co-moving distance to
the E+A galaxy at redshift $z$ and $\Delta\theta$ is the angular
separation of the galaxy and cluster centroid; and the radial
(redshift-space) physical distance $D_r = (c \, \Delta z/H)/(1+z)$,
where $H$ is the Hubble constant at the E+A galaxy redshift and
$\Delta z$ is the redshift separation of the galaxy and cluster
centroid.  The radial separation was considered independently because
its apparent value can be significantly enhanced by line-of-sight
peculiar velocities.  An E+A galaxy was considered a `cluster' object
if both $D_t < r_0$ and $D_r < \sqrt{(r_0)^2 + (2\sigma/H)^2}$, where
$\sigma$ is the velocity dispersion of the cluster in question,
typically $\sigma \sim 1000$ km s$^{-1}$.  The selection volume is
thus broadened in the radial direction to allow for peculiar
velocities.  The critical physical scale $r_0$ was taken as 5 Mpc
(chosen to exceed comfortably the virial radius of a typical rich
cluster).

We restricted our cluster analysis to the SGP region of the galaxy
survey; the NGP region is only covered by the Abell catalogue thus the
catalogued cluster distribution is much sparser.  Using the
average-Balmer (or H$\delta$) E+A catalogue: of 37 (122) galaxies in
the SGP region, 4 (20) were classified as cluster objects using the
method described above.  {\it We conclude that the 2dFGRS E+A galaxies
typically lie in the field, outside clusters, in agreement with the
findings of Zabludoff et al.\ (1996).}

\subsection{Cross-correlation with remaining 2dFGRS galaxies}

As a second means of quantifying the environments of the E+A galaxy
samples, we counted the number of objects in the 2dFGRS within a
co-moving radius $r_0 = 8 \, h^{-1}$ Mpc of each E+A galaxy, making no
attempt to correct for peculiar velocities.  By integrating the
measured 2dFGRS $\bj$-band luminosity function (Norberg et al.\ 2002)
to the survey apparent magnitude limit, we can determine the average
number of objects expected within this sphere at any given redshift
(in the absence of clustering) and hence obtain a local overdensity
(e.g.\ on a scale $r_0 = 8 \, h^{-1}$ Mpc) for each E+A galaxy.  We
then compared the average local overdensity of E+A galaxies with that
of a sample of randomly-drawn 2dFGRS objects, for which we applied an
identical technique.

Using the average-Balmer (or H$\delta$) E+A catalogue, the average
overdensity $\delta$ within a sphere of co-moving radius $r_0 = 8 \,
h^{-1}$ Mpc centred on an E+A galaxy was $\delta = 0.96 \pm 0.27$
($1.03 \pm 0.15$).  The quoted error was derived as the error in the
mean of the overdensity distribution, $\sigma/\sqrt{N}$, where
$\sigma$ is the standard deviation of the distribution and $N$ is the
number of E+A galaxies in the sample.  Measuring the average
overdensity of 10000 randomly-drawn 2dFGRS catalogue members by an
identical method yielded $\delta = 1.18$ (with a negligible error).

Motivated by the spheroidal morphologies of many of the average-Balmer
E+A galaxies (Section \ref{secmorph}), we also defined a `control
sample' of 2dFGRS `elliptical' galaxies.  These were selected based on
the classification system developed for 2dFGRS spectra using
principal-component analysis (Madgwick et al.\ 2002).  This procedure
assigns a parameter $\eta$ to each spectrum based upon the strength of
nebular emission; the value of this continuous variable $\eta$ turns
out to correlate relatively well with the galaxy $\bj$-band morphology
(Madgwick 2003).  Following Magdwick et al.\ (2002) we define a
`relatively quiescent' (i.e.\ early-type) sub-sample by the cut $\eta
< -1.4$.  We found that the average overdensity of 10000 2dFGRS
`ellipticals', selected in this manner, to be $\delta = 1.22$.

\begin{figure}
\center
\epsfig{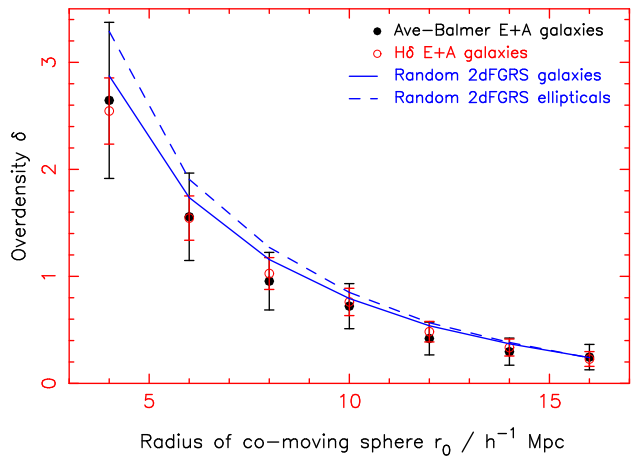}
\caption{The overdensity at the locations of the E+A galaxies,
obtained by counting the number of 2dFGRS catalogue entries within a
sphere of co-moving radius $r_0$ and comparing the result to that
obtained by integrating the 2dFGRS luminosity function to the survey
apparent magnitude limit.  A similar analysis was performed for 10000
randomly-drawn 2dFGRS catalogue members (solid line) and the same
number of 2dFGRS `ellipticals' (dashed line), the latter sample
selected by requiring the spectral classification parameter $\eta <
-1.4$.}
\label{figoverdens}
\end{figure}

We repeated this analysis for other co-moving radii $r_0$ (Fig.\
\ref{figoverdens}), finding that on all scales {\it the mean
overdensity at the locations of the E+A galaxies is statistically
consistent with the ensemble of 2dFGRS galaxies}.  This result is in
agreement with the findings of Quintero et al.\ (2004) based on SDSS
spectra containing an excess signature of A stars relative to K stars.
Furthermore, the overdensity pattern around elliptical galaxies is not
favoured as a description of E+A galaxy environments, even though many
average-Balmer E+A systems possess spheroidal morphologies.  As
illustrated by Fig.\ \ref{figoverdens}, we recover the well-known
result that elliptical galaxies inhabit denser local environments than
the ensemble of galaxies.

\subsection{Cross-correlation with 2dFGRS galaxy groups catalogue}

Eke et al.\ (2004a) have constructed a catalogue of galaxy groups from
the 2dFGRS, by means of a friends-of-friends percolation algorithm.
The data are publicly available online at {\tt
http://www.mso.anu.edu.au/2dFGRS/Public/2PIGG/} and consist of two
tables.  For each 2dFGRS galaxy, the first table lists the group ID or
a flag indicating that the galaxy was ungrouped.  For each group, the
second table records the number of group members and details such as
the projected linear size and velocity dispersion of the group.

For each galaxy in our E+A catalogues, we first considered the number
of members of the group in which that galaxy resides.  The
distributions of group membership numbers for E+A galaxies and for
randomly-drawn 2dFGRS galaxies were statistically consistent.  As
such, {\it approximately 50 per cent of our E+A galaxies are
isolated}, not linked to any companion by the percolation algorithm.

\begin{figure}
\center
\epsfig{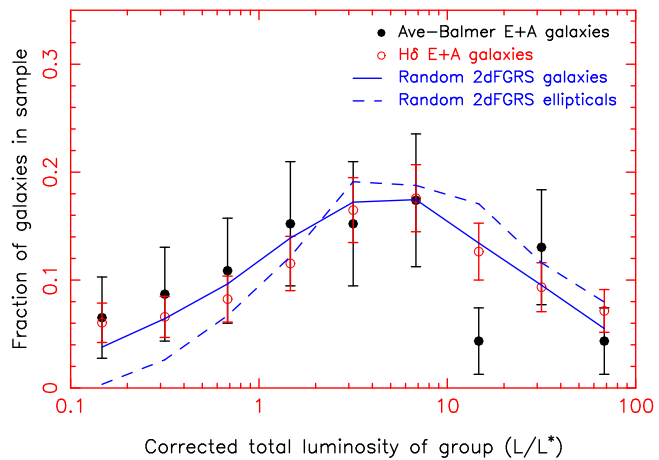}
\caption{The distribution of group luminosities for the average-Balmer
(solid circles) and H$\delta$ (open circles) E+A galaxy catalogues.
The luminosities are corrected by a factor compensating for those
galaxies falling beneath the apparent magnitude threshold at the group
redshift.  The error in the number of objects $N$ in each bin was
taken as the Poisson error $\sqrt{N}$.  The solid and dashed lines
represent the distribution of group luminosities obtained for 1000
randomly-drawn 2dFGRS galaxies and ellipticals, respectively.}
\label{figgrouplum}
\end{figure}

However, group membership is a poor indicator of group size: massive
groups can have low membership simply because they are at high
redshift.  A better quantity to consider is the corrected total
luminosity of the group (see Eke et al.\ 2004b).  This is the summed,
weighted luminosities of the galaxies in the group, divided by an
incompleteness factor compensating for those galaxies falling beneath
the apparent magnitude threshold at the group redshift.  Fig.\
\ref{figgrouplum} plots the distribution of these group luminosities
for the E+A catalogues.  We overplot the results of an identical
analysis for samples of randomly-drawn 2dFGRS galaxies and
ellipticals: the latter inhabit preferentially more luminous groups.
{\it We find no evidence that E+A galaxies inhabit groups of a
different nature to the overall galaxy population.}  Furthermore, the
groups containing E+A galaxies appear dissimilar to those hosting
elliptical galaxies; the elliptical model is rejected by the
average-Balmer (or H$\delta$) data with a confidence of $99.7\%$
($98.0\%$), based on the value of the chi-squared statistic.

\subsection{Cross-correlation with Supercosmos Sky Survey catalogues}
\label{seclocenv}

As a final method of investigating the environment of 2dFGRS E+A
galaxies, we obtained the Supercosmos Sky Survey (SSS) photometric
catalogues for the regions surrounding each E+A galaxy.  The
approximate magnitude limits of the photographic plates which are
scanned to produce the SSS are $\bj = 22.5$ and $\rf = 21.5$.  We
downloaded SSS object catalogues using the web interface {\tt
http://www-wfau.roe.ac.uk/sss/obj\_batch\_email.html} specifying a
circular extraction with radius 5 arcmin.

Using the 2dFGRS $\bj$-band luminosity function (Norberg et al.\
2002), we can deduce the apparent magnitude $\bj^*(z)$ corresponding
to an absolute magnitude $M_b^*$ at the redshift $z$ of a sample E+A
galaxy.  We then define a `bright neighbour' as a nearby SSS galaxy
with $\bj < \bj^*(z) + 1$.  We define a `faint neighbour' as a nearby
SSS galaxy with $\bj^*(z) + 1 < \bj < 22.5$.  The presence of a nearby
`bright' or `faint' neighbour may indicate, respectively, an imminent
major or minor merger.

\begin{figure}
\center
\epsfig{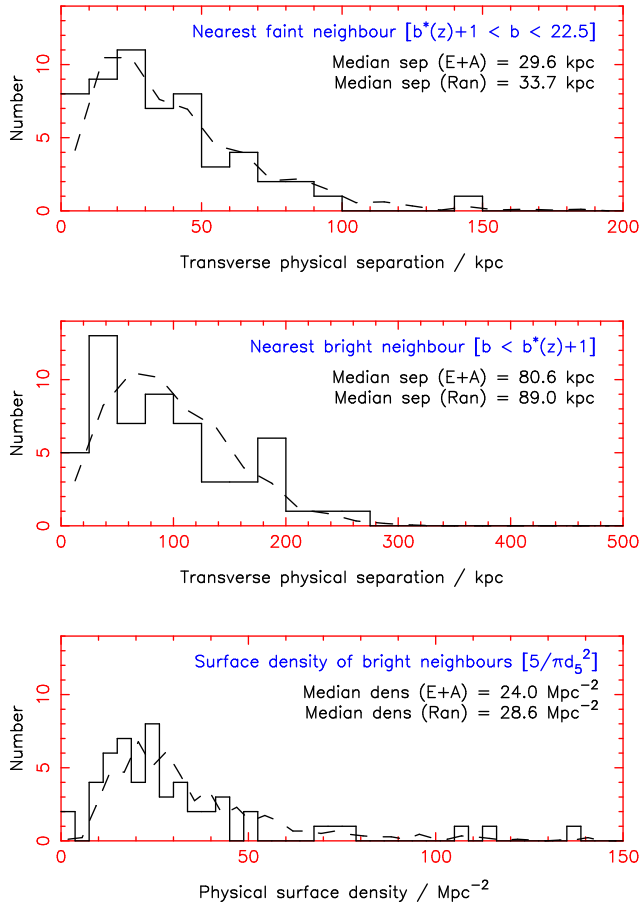}
\caption{The solid histograms plot the distribution of local
environments of the average-Balmer E+A galaxy sample (56 objects),
derived by overlaying the E+A galaxy positions on the Supercosmos Sky
Survey (SSS) catalogue.  The panels illustrate the local environment
quantified by the nearest faint neighbour, the nearest bright
neighbour and the local surface density of bright neighbours.  The
dashed line displays the same statistics determined for 1000
randomly-selected 2dFGRS galaxies.  See Section \ref{seclocenv} for
more details.}
\label{figlocab}
\end{figure}

\begin{figure}
\center
\epsfig{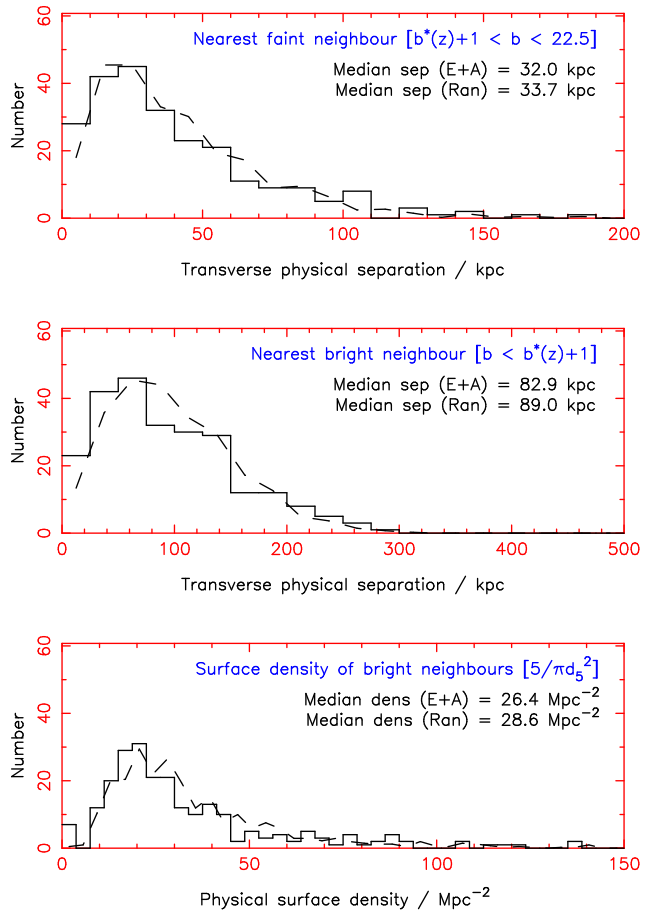}
\caption{The same statistics as Fig.\ \ref{figlocab}, determined for
the H$\delta$ E+A galaxy sample (243 objects).}
\label{figlochd}
\end{figure}

Using the known E+A galaxy redshift we can convert angular separations
into transverse physical separations.  For each E+A galaxy we derived:
\begin{itemize}
\item The transverse physical separation (in kpc) of the nearest faint
neighbour.
\item The transverse physical separation (in kpc) of the nearest
bright neighbour.
\item The physical surface density (in Mpc$^{-2}$) defined by the five
nearest bright neighbours ($\Sigma = 5/\pi d_5^2$, where $d_5$ is the
transverse physical separation of the fifth nearest bright neighbour).
\end{itemize}
Note that each of these quantities is insensitive to the effects of
peculiar velocities.  We compared these statistics to those determined
for a sample of 1000 randomly-drawn 2dFGRS galaxies (restricted to
redshifts $0.002 < z < 0.25$ to eliminate stellar contamination).  The
results for the two E+A galaxy catalogues are displayed in Figs.\
\ref{figlocab} and \ref{figlochd}, with the results for the random
sample overplotted in each case.

We used the Kolmogorov-Smirnov (K-S) statistical test to ascertain
whether the local environments of the E+A galaxy samples were drawn
from a different parent distribution to the local environments of the
random 2dFGRS sample.  Let $p$ be the probability that the K-S
statistic could exceed the observed value by random chance, if the two
distributions were drawn from the same parent distribution.  The
values of $p$ for the distributions of nearest faint neighbour,
nearest bright neighbour and local surface density, for the
average-Balmer (H$\delta$) E+A catalogue, were $0.58$ ($0.35$), $0.32$
($0.11$) and $0.21$ ($0.06$).  {\it We conclude that there is no
significant evidence that the distribution of E+A galaxy local
environments is different from that of the 2dFGRS sample as a whole.}

\section{Clustering properties}
\label{secclus}

We quantified the clustering properties of 2dFGRS E+A galaxies using a
spatial correlation function analysis.  Such an analysis yields
information about the bias of these galaxies with respect to the
underlying matter density field, which we can compare to the bias of
other classes of galaxy.  In the approximation that E+A galaxies
possess a linear bias $b_e$, their spatial auto-correlation function
$\xi_{ee}$ as a function of spatial separation $r$ takes the form
\begin{equation}
\xi_{ee}(r) = b_e^2 \, \xi_{mm}(r) \: ,
\end{equation}
where $\xi_{mm}$ is the spatial auto-correlation function of the
underlying matter density field.

As our samples contain insufficient E+A galaxies to perform an
auto-correlation function analysis (there is only one pair of objects
in the average-Balmer catalogue with a redshift-space separation less
than $20 \, h^{-1}$ Mpc), we instead determined the cross-correlation
function $\xi_{eg}$ of the E+A galaxy samples with the rest of the
2dFGRS catalogue.  In the approximation of linear bias,
\begin{equation}
\xi_{eg}(r) = b_e \, b_g \, \xi_{mm}(r) \: ,
\end{equation}
where $b_g$ is the bias factor for an average 2dFGRS galaxy.  If we
additionally measure the spatial auto-correlation function of 2dFGRS
galaxies, $\xi_{gg}$, we can estimate the relative bias of E+A
galaxies as
\begin{equation}
\frac{b_e}{b_g} = \frac{\xi_{eg}(r)}{\xi_{gg}(r)} \: .
\end{equation}

We measured all correlation functions in redshift space, making no
attempt to correct for peculiar velocities.  The cross-correlation
function is measured by comparing the cross-pair counts of the E+A
sample with, respectively, the full 2dFGRS catalogue (containing $n_g$
galaxies) and a random distribution of $n_r$ points possessing the
same selection function as the full catalogue.  The random
distributions were generated using the publicly-available 2dFGRS mask
software written by Peder Norberg and Shaun Cole (see {\tt
http://msowww.anu.edu.au/2dFGRS/}).  Denoting these respective
cross-pair counts as $N_{eg}(s)$ and $N_{er}(s)$, with $s$ denoting a
redshift-space separation, we used the standard estimator for the
cross-correlation function $\xi_{eg}$:
\begin{equation}
\xi_{eg}(s) = \frac{n_r}{n_g} \frac{N_{eg}(s)}{N_{er}(s)} - 1 \: .
\label{eqcrosscorr}
\end{equation}
When determining the auto-correlation function of 2dFGRS galaxies
$\xi_{gg}$ for comparison, we used the standard estimator equivalent
to equation \ref{eqcrosscorr}:
\begin{equation}
\xi_{gg}(s) = \frac{2 \, n_r}{n_g} \frac{N_{gg}(s)}{N_{gr}(s)} - 1 \:
.
\label{eqautocorr}
\end{equation}
The extra factor of $2$ in equation \ref{eqautocorr} arises because
$N_{gg}$ is a auto-pair count rather than a cross-pair count (thus
contains half the number of unique pairs).

\begin{figure}
\center
\epsfig{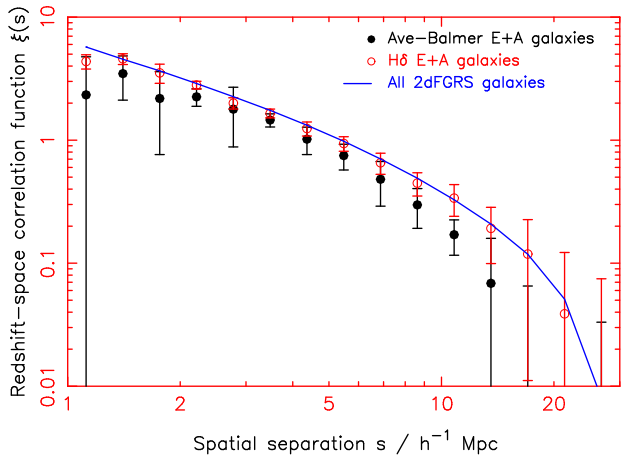}
\caption{Spatial correlation function measurements for the E+A galaxy
samples.  The solid and open circles plot the cross-correlation
function of respectively the average-Balmer and the H$\delta$ E+A
galaxies with the remainder of the 2dFGRS catalogue, measured using
equation \ref{eqcrosscorr}.  The error bar for each separation bin is
estimated using the jack-knife re-sampling technique, as described in
the text.  The solid line is the auto-correlation function of the
entire 2dFGRS catalogue, derived employing equation \ref{eqautocorr}.
The error in this function (not shown) is negligible compared to the
error in the other measurements.}
\label{figcorr}
\end{figure}

We included the 2dFGRS NGP and SGP regions, but not the random fields,
in our clustering analysis.  The total numbers of galaxies analyzed
were 50, 201 and 195188 for the average-Balmer catalogue, the
H$\delta$ catalogue and the entire 2dFGRS, respectively.  The results
are displayed in Fig.\ \ref{figcorr}.  Poisson error bars are often
assigned to correlation function measurements, but these are known to
underestimate the true variance of the estimators of equations
\ref{eqcrosscorr} and \ref{eqautocorr} by a significant factor (e.g.\
Landy \& Szalay 1993).  Instead, we estimated a realistic statistical
variance using the `jack-knife' approach.  We divided both the NGP and
SGP strips into four quadrants and repeated the correlation function
estimation 8 times, keeping 7 quadrants; the error for each separation
bin is then estimated by multiplying by $\sqrt{8}$ the resulting
standard deviation across the eight sub-samples.

The correlation function estimate for the H$\delta$ sample agrees well
with that for the entire 2dFGRS (implying $b_e \approx b_g$) but there
is evidence that the average-Balmer catalogue is somewhat less
clustered.  Considering the small size of this sample (50 objects) and
the strong statistical correlation between neighbouring bins of the
$\xi(r)$ estimator, we consider this result to be tentative.  On
scales where $\xi \ll 1$, systematic errors probably dominate the
uncertainities in the correlation functions.  In particular, there is
a well-known systematic effect (the `integral constraint') caused by
the uncertainty in estimating the mean density of the galaxy sample.
This effect distorts the measured values of $\xi$ systematically
downwards on large scales (in a similar way for all jack-knife
re-samples).

\section{Luminosity function}
\label{seclf}

The distribution of luminosities of E+A galaxies may encode
information about the physical mechanisms responsible for their
formation.  If, for example, E+A galaxies possess below-average
luminosities, then they would be more susceptible to tidal
interactions, which would then be implicated as a formation mechanism.

We determined the $\bj$-band luminosity function, $\phi(M)$, of the
two samples of E+A galaxies using the step-wise maximum likelihood
(SWML) method (Efstathiou, Ellis \& Peterson 1988).  We also measured
the $\bj$-band luminosity function of the full 2dFGRS catalogue and
the ellipticals sample (see Section \ref{secenv}), using an identical
method.  In order to limit the effects of incompletenesses, we
restricted our analysis to the apparent magnitude range $14 < \bj <
19.2$.  The total numbers of galaxies analyzed were respectively 53,
174, 166243 and 60640 for the average-Balmer catalogue, the H$\delta$
catalogue, the entire 2dFGRS and the ellipticals sample.  When
evaluating the luminosity function we used the Supercosmos Sky Survey
magnitudes, together with the K-correction described in Section
\ref{seccol} combined with an `E-correction' for the luminosity
evolution of an average 2dFGRS galaxy (Norberg et al.\ 2002).  The
fractional error in the measurement of $\phi(M)$ in each luminosity
band is taken as a Poisson error $1/\sqrt{N}$, where $N$ is the number
of objects falling in that band.

\begin{figure}
\center
\epsfig{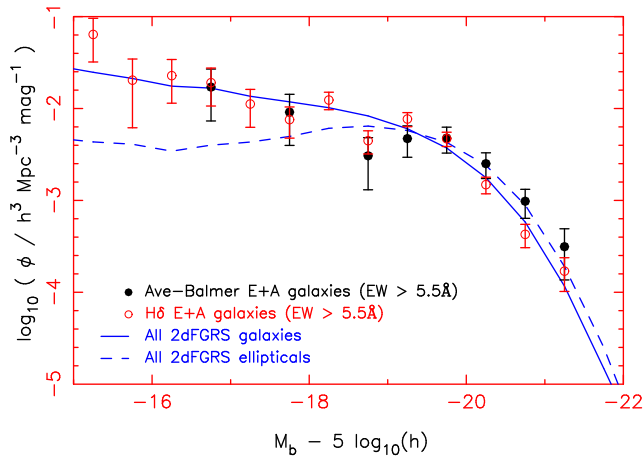}
\caption{The luminosity functions of the average-Balmer E+A galaxy
sample (53 objects, solid circles), the H$\delta$ E+A sample (174
objects, open circles), the whole 2dFGRS catalogue (166243 objects,
solid line) and the ellipticals sample (60640 objects, dashed line).
All measurements have been derived using a step-wise maximum
likelihood code and normalized to a consistent source surface density
$\sigma = 141.4$ deg$^{-2}$.  The errors in the luminosity function
measurements for the 2dFGRS catalogues are not much greater than the
thickness of the lines and are omitted for clarity.}
\label{figlf5pt5}
\end{figure}

The SWML estimator is unnormalized; we normalized each luminosity
function to an equivalent source surface density $\sigma$ which we
fixed as follows.  In the absolute magnitude interval $-16.5 > M - 5
\, {\rm log_{10}} \, h > -22$ the 2dFGRS luminosity function is
accurately described by a Schechter function with parameters $M^* - 5
\, {\rm log_{10}} \, h = -19.66$, $\alpha = -1.21$ and $\Phi^* = 1.61
\times 10^{-2} \, h^3$ Mpc$^{-3}$ (Norberg et al.\ 2002).  We obtained
a normalization $\sigma = 141.4$ deg$^{-2}$ by integrating this
Schechter function over the redshift interval $0.002 < z < 0.3$ and
the apparent magnitude interval $14 < \bj < 19.2$.  By applying a
consistent normalization in this manner, we are able to investigate
any difference of shape between the luminosity functions for the
2dFGRS and for the E+A galaxy samples.  The results are displayed in
Fig.\ \ref{figlf5pt5}.

\begin{figure}
\center
\epsfig{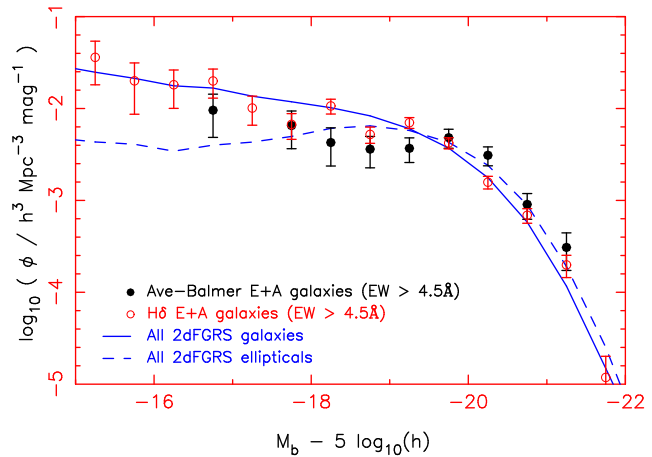}
\caption{The luminosity functions of extended E+A galaxy catalogues,
for which the absorption equivalent width threshold has been reduced
from $5.5 \ang$ to $4.5 \ang$.  The numbers of average-Balmer and
H$\delta$ E+A galaxies analyzed in this plot were respectively 87 and
292.  The results are displayed in the same style as Fig.\
\ref{figlf5pt5}.}
\label{figlf4pt5}
\end{figure}

At brighter luminosities ($M - 5 \, {\rm log_{10}} \, h < -20$), Fig.\
\ref{figlf5pt5} hints that the average-Balmer sample may possess a
luminosity distribution more consistent with that of elliptical
galaxies than average 2dFGRS galaxies (although formally, the
luminosity functions of both E+A catalogues are statistically
consistent with the whole 2dFGRS database; the chi-squared
probabilities in the two cases are $0.09$ and $0.06$, considering bins
containing $\ge 2$ objects).  Investigating further, we lowered the
absorption equivalent width threshold for E+A catalogue selection from
$5.5 \ang$ to $4.5 \ang$, in order to create larger samples.  The
luminosity functions of these extended catalogues are displayed in
Fig.\ \ref{figlf4pt5}.  The luminosity distribution of the
average-Balmer E+A galaxies is now {\it inconsistent} with that of the
overall 2dFGRS catalogue (with significance $> 99.99\%$) and
consistent with the luminosity function of ellipticals (the
chi-squared probability is $0.19$).  The result for the
H$\delta$-selected sample remains consistent with that for the
ensemble of 2dFGRS galaxies.  This finding seems very reasonable given
the morphological distributions determined in Section \ref{secmorph}:
the average-Balmer catalogue is dominated by spheroidal galaxies,
whereas the H$\delta$ sample contains a significant admixture of disk
galaxies.

We conclude that the distribution of luminosities of
H$\delta$-selected galaxies is consistent with the entire 2dFGRS
population, but there is evidence to suggest that the luminosity
function of the average-Balmer sample better matches that of
(spectroscopically-defined) `elliptical' galaxies.

We repeated the statistical analyses of local environment (Section
\ref{secenv}) and clustering (Section \ref{secclus}) for the lower
($4.5 \ang$) absorption equivalent width threshold, and found that the
conclusions remain unchanged.

\section{Discussion}

We have selected samples of low-redshift E+A galaxies from 161000
high-quality spectra in the 2dFGRS, constituting a large-scale
environmentally-unbiased study of E+A systems in the local Universe.
We used two different selection techniques: the first utilizing three
Balmer absorption lines (H$\delta$, H$\gamma$, H$\beta$) together with
the [OII] $3727 \ang$ feature, and the second technique employing
solely the H$\delta$ and [OII] lines.  These methods resulted in the
selection of 56 and 243 galaxies, respectively.

We inspected the morphologies of the E+A galaxies using images drawn
from the Supercosmos Sky Survey (SSS), finding that:
\begin{itemize}
\item The largest sub-population of `average-Balmer' E+A galaxies is
E/S0 systems.  There are no disk-dominated galaxies.  In contrast, the
largest sub-sample of H$\delta$-selected objects is late-type spirals.
\item In a small but significant number of cases, the SSS images
reveal evidence of recent major galaxy mergers, such as disturbed
morphologies, coalescing disks and tidal tails or envelopes.  We
detect a notable (50 kpc) tidal tail associated with one object.
\item Further study (i.e.\ spatially-resolved spectroscopy) is
necessary to establish the influence of {\it aperture effects} due to
the $2''$ angular diameter of the 2dF spectrograph fibres being
significantly smaller than the angular size of most of the galaxies.
\end{itemize}

The individual catalogues resulting from the two E+A selection
techniques contain a different distribution of galaxy populations.
Less than $10\%$ of average-Balmer E+A galaxies yield any detection of
H$\alpha$ emission, with the highest deduced star-formation rate being
$0.17 \, M_\odot$ yr$^{-1}$.  In contrast, $60\%$ of spectra selected
by strong H$\delta$ absorption (and no [OII] emission) contain a
measurable H$\alpha$ line (although only rarely does the inferred
star-formation rate exceed $1 \, M_\odot$ yr$^{-1}$).

These samples permitted a statistical investigation of the
environments and luminosities of low-redshift E+A galaxies.  We
compared the two E+A catalogues to both the entire 2dFGRS database and
a sub-population of 2dFGRS `ellipticals', the latter selected using
the spectral classification parameter $\eta$.  With regard to the
environments of E+A galaxies, we found that:
\begin{itemize}
\item E+A galaxies at low redshift lie predominantly in the field
rather than in clusters, in agreement with the analysis of Zabludoff
et al.\ (1996).
\item Cross-matching E+A galaxies with the 2dFGRS galaxy groups
catalogue of Eke et al.\ (2004), we determined that the distribution
of membership and luminosity of the groups in which E+A systems reside
was consistent with that of the overall galaxy population.  In
particular, approximately 50 per cent of E+A objects are classified as
isolated galaxies.  The distribution of groups containing E+A systems
is dissimilar to those hosting elliptical galaxies.
\item Nearest-neighbour and correlation-function analyses confirmed
the conclusion that {\it the local environments of E+A galaxies are
consistent with those of the ensemble of 2dFGRS galaxies.}
\item These conclusions hold true for both E+A galaxy catalogues.
\end{itemize}

Concerning the luminosity function of E+A galaxies:
\begin{itemize}
\item Average-Balmer E+A galaxies possess a luminosity distribution
matching that of elliptical galaxies, and disagreeing with the overall
2dFGRS luminosity function (although this result only became
statistically significant when the equivalent width selection
threshold was reduced from $5.5 \ang$ to $4.5 \ang$).
\item The luminosity distribution of H$\delta$ E+A galaxies matches
that of the full 2dFGRS population.
\end{itemize}

What can we conclude from these results?  Considering first the
average-Balmer E+A catalogue: the preference for spheroidal
morphologies, the incidence of identifiable merger remnants, and the
match of the luminosity distribution to that of elliptical galaxies is
all consistent with {\it major galaxy mergers} being an important
formation process for these galaxies.  Indeed, this model has proven
successful in theoretical simulations (Bekki et al.\ 2001).  Thus it
is interesting that the distribution of local environments of these
E+A systems matches that of the ensemble of galaxies; for example,
there is no correlation with the properties of galaxy groups.  This in
part reflects the fact that at low redshift the merger probability is
enhanced in the field with respect to clusters.  Elliptical galaxies
forming today do not follow the classic morphology-density relation,
because the relative velocities of galaxies in denser environments is
too high to permit mergers: cluster sub-structure has become
dynamically relaxed.

The impressive manner in which the distribution of local environments
of E+A systems traces that of the ensemble of 2dFGRS galaxies implies
that the E+A galaxy formation mechanism is driven by very local
encounters, without reference to the wider group environment.
Theoretical simulations indicate that an E+A spectrum marks a late
stage of a merger, when the cores of the merging systems have
coalesced and the merging companion is no longer identifiable (Bekki
et al.\ 2001): we would not expect a surplus of close companions.
Furthermore, we note a consistency of our findings with the dependence
of star formation upon environment (e.g.\ Balogh et al.\ 2004).  In
such studies, the only environmentally-selected population at low
redshift to show enhancements of star formation are close galaxy
pairs, which are no longer distinguishable at the onset of the E+A
phase.

Turning now to galaxies selected by H$\delta$ absorption alone: the
greater incidence of disk-like morphologies and detectable H$\alpha$
emission demonstrates that star formation is still ongoing in many of
these objects, albeit at a relatively low level ($< 1 \, M_\odot$
yr$^{-1}$).  Such systems therefore cannot be considered `true E+A
galaxies'.  In these cases, [OII] emission may either be suppressed by
dust obscuration or present with a low signal-to-noise ratio, and the
H$\gamma$ and H$\beta$ absorption features are presumably subject to
emission-filling.

Our E+A galaxy catalogues represent a useful database for follow-up
studies.  In particular, spatially-resolved spectroscopy is a critical
probe of the formation mechanism (e.g.\ Norton et al.\ 2001), mapping
the kinematics of the stellar populations and the star-formation
history of the galaxies as a function of position.  In addition,
high-resolution optical imaging can define the morphologies with
greater fidelity, and imaging in 21 cm neutral hydrogen emission
yields the large-scale distribution and kinematics of remaining gas,
encoding information about any merger event.

\section*{Acknowledgments}

We thank Elaine Sadler for useful conversations about this work.  CB,
WJC and KB acknowledge the financial support of the Australian
Research Council throughout the course of this work.  MBP was
supported by an Australian Postgraduate Award.

\pagebreak
\begin{table*}
\center
\caption{The 2dFGRS average-Balmer E+A galaxy catalogue (56 members).
Photographic $\bj$ and $\rf$ magnitudes were obtained from the
Supercosmos Sky Survey; $\bj$ was converted to an absolute magnitude
$M_b$ using the K-correction described in Section \ref{seccol}.
Balmer equivalent widths (in $\ang$) originate from the 2dFGRS line
fits catalogue and are quoted as rest-frame values.}
\label{tabavebalm}
\begin{tabular}{ccccccccccc}
\hline
Serial No. & Name & $\alpha$ (J2000) & $\delta$ (J2000) & $z$ & $\bj$
& $\rf$ & $M_b$ & H$\delta$ & H$\gamma$ & H$\beta$ \\
\hline
   218 & TGS495Z048 &  0 11 22.88 & -33 25  3.3 & 0.100 & 18.53 & 17.24 & -19.29 &  -5.33 &  -5.64 &  -5.93 \\
  1232 & TGS431Z066 & 23 55 41.83 & -32 21 32.7 & 0.156 & 19.19 & 17.92 & -19.81 &  -5.35 &  -4.28 &  -6.42 \\
  6549 & TGS439Z075 &  0 29 10.97 & -32 42 34.2 & 0.108 & 17.89 & 16.61 & -20.12 &  -7.65 &  -5.43 &  -7.53 \\
  6833 & TGS438Z051 &  0 25 14.18 & -32 15  1.8 & 0.134 & 18.35 & 16.90 & -20.32 &  -1.37 &  -1.66 & -12.75 \\
  7402 & TGS497Z208 &  0 18 29.19 & -33 34 55.3 & 0.049 & 18.94 & 18.30 & -16.98 &  -7.70 &  -6.17 &  -6.43 \\
  8648 & TGS555Z002 &  0 24 11.03 & -34 54 52.1 & 0.240 & 18.87 & 17.49 & -21.39 &  -5.47 &  -5.72 &  -4.92 \\
 11882 & TGS502Z221 &  0 51 31.48 & -32 48 50.5 & 0.111 & 17.75 & 16.51 & -20.33 &  -5.01 &  -4.41 &  -6.64 \\
 30347 & TGS519Z127 &  2 35 47.21 & -33 35 15.6 & 0.078 & 18.52 & 17.36 & -18.63 &  -6.05 &  -5.21 &  -6.38 \\
 30561 & TGS519Z227 &  2 33 10.60 & -33 52 24.4 & 0.070 & 17.64 & 16.30 & -19.30 &  -6.21 &  -7.10 &  -6.34 \\
 32941 & TGS574Z157 &  2 18 17.27 & -35 27 27.7 & 0.223 & 18.99 & 17.80 & -20.93 &  -5.24 &  -5.11 &  -6.22 \\
 35418 & TGS520Z261 &  2 40 24.27 & -33 25 50.6 & 0.035 & 18.17 & 17.15 & -17.08 &  -6.34 &  -6.68 &  -5.41 \\
 46013 & TGS480Z208 & 22 18 22.99 & -33  2 36.7 & 0.101 & 18.28 & 17.01 & -19.56 &  -3.82 &  -4.95 &  -7.08 \\
 48663 & TGS589Z117 & 22 22 52.22 & -36 57  1.3 & 0.033 & 15.46 & 14.48 & -19.63 &  -7.44 &  -8.47 &  -7.69 \\
 55742 & TGS541Z024 & 23 25 58.09 & -33 56 59.9 & 0.063 & 18.60 & 17.51 & -18.00 &  -5.93 &  -4.62 &  -5.98 \\
 56491 & TGS539Z123 & 23 15 25.13 & -35 12 59.1 & 0.196 & 18.39 & 17.09 & -21.25 &  -6.69 &  -6.80 &  -6.39 \\
 63733 & TGS278Z037 &  0 11  9.90 & -28 47 40.0 & 0.070 & 19.26 & 18.51 & -17.50 &  -6.34 &  -5.78 &  -5.62 \\
 66097 & TGS359Z193 &  0  4 13.85 & -30 19 49.9 & 0.221 & 18.91 & 17.51 & -21.13 &  -6.62 &  -6.04 &  -7.01 \\
 66348 & TGS358Z179 & 23 59 29.87 & -30 16 21.9 & 0.120 & 18.06 & 16.73 & -20.25 &  -5.41 &  -5.80 &  -6.03 \\
 72301 & TGS438Z207 &  0 25  1.82 & -31  1 21.1 & 0.128 & 18.92 & 17.58 & -19.58 &  -3.71 &  -5.77 &  -5.93 \\
 77063 & TGS444Z180 &  0 55 49.58 & -31 47 44.9 & 0.179 & 19.01 & 17.52 & -20.51 &  -5.17 &  -5.72 &  -5.70 \\
 78421 & TGS294Z049 &  1 14 55.37 & -28 55 52.1 & 0.186 & 19.16 & 17.78 & -20.39 &  -8.18 &  -7.98 &  -7.30 \\
 80795 & TGS373Z107 &  1  5 41.72 & -29 36 55.2 & 0.195 & 18.55 & 17.14 & -21.15 &  -7.72 &  -6.73 &  -5.35 \\
 91037 & TGS231Z063 &  2 29 44.49 & -27 54 59.4 & 0.198 & 19.01 & 17.78 & -20.61 &  -7.17 &  -7.00 &  -7.99 \\
 93583 & TGS387Z032 &  2 15 40.52 & -30 50 54.5 & 0.092 & 17.69 & 16.53 & -19.88 &  -6.99 &  -5.13 &  -6.02 \\
 98002 & TGS464Z179 &  2 36 17.68 & -31 31 36.8 & 0.165 & 18.93 & 17.74 & -20.18 &  -6.26 &  -6.08 &  -6.75 \\
103972 & TGS402Z315 &  3 31  7.88 & -30 35 54.0 & 0.066 & 18.98 & 17.67 & -17.78 &  -4.40 &  -4.19 &  -7.10 \\
104025 & TGS322Z289 &  3 30 32.41 & -29 18 48.8 & 0.120 & 18.32 & 17.03 & -19.98 &  -6.84 &  -7.49 &  -7.56 \\
108920 & TGS246Z107 & 21 46 15.20 & -29 43 43.9 & 0.021 & 17.06 & 16.08 & -16.99 &  -7.09 &  -5.74 &  -5.32 \\
115650 & TGS259Z083 & 22 42 28.76 & -28 35  4.5 & 0.028 & 15.51 & 14.52 & -19.24 &  -6.50 &  -7.03 &  -6.01 \\
116268 & TGS343Z007 & 22 50 58.49 & -30 17 32.4 & 0.202 & 18.71 & 17.36 & -21.04 &  -5.86 &  -5.90 &  -4.99 \\
118891 & TGS266Z090 & 23 10 46.57 & -28 31 49.7 & 0.088 & 17.73 & 16.57 & -19.71 &  -6.04 &  -5.50 &  -5.34 \\
124944 & TGS350Z150 & 23 26 36.76 & -30 19 27.4 & 0.158 & 18.37 & 17.07 & -20.68 &  -6.18 &  -6.15 &  -5.95 \\
127591 & TGS195Z088 & 23 48 21.17 & -27 57 32.5 & 0.177 & 19.23 & 17.79 & -20.22 &  -7.74 &  -6.12 &  -3.37 \\
129207 & TGS271Z130 & 23 41  8.90 & -28 55 25.4 & 0.082 & 18.30 & 17.08 & -18.99 &  -4.96 &  -4.88 &  -6.88 \\
164029 & TGS109Z228 &  3  4 49.78 & -24 57 57.1 & 0.191 & 19.31 & 18.15 & -20.17 &  -8.96 &  -6.48 &  -5.53 \\
182340 & TGS120Z302 & 22 37 37.03 & -25 50 57.6 & 0.235 & 18.45 & 17.39 & -21.50 &  -6.26 &  -6.37 &  -5.66 \\
182382 & TGS120Z325 & 22 36  6.92 & -26 18 52.9 & 0.005 & 15.78 & 14.69 & -15.03 &  -9.38 &  -7.73 &  -6.94 \\
194184 & TGN154Z163 & 10  4 24.87 &  -4 25 27.0 & 0.129 & 18.27 & 17.05 & -20.19 &  -5.61 &  -6.18 &  -5.64 \\
208265 & TGN224Z107 & 10 33 31.22 &  -3 20 36.0 & 0.098 & 18.55 & 17.47 & -19.13 &  -6.18 &  -5.88 &  -5.74 \\
215246 & TGN169Z023 & 11 25 26.17 &  -4 32 36.8 & 0.082 & 17.92 & 16.76 & -19.34 &  -6.53 &  -5.07 &  -6.31 \\
215346 & TGN169Z051 & 11 24 55.68 &  -3 47 35.7 & 0.017 & 17.43 & 16.73 & -16.21 &  -4.56 &  -5.04 &  -6.50 \\
241469 & TGN192Z151 & 13  0 16.53 &  -2 54 33.3 & 0.102 & 18.73 & 17.69 & -19.03 &  -6.28 &  -5.57 &  -5.44 \\
280809 & TGN369Z015 & 11 11  8.13 &  +0 40 48.5 & 0.185 & 18.53 & 17.33 & -20.88 &  -7.44 &  -6.65 &  -8.51 \\
287603 & TGN441Z175 & 11 23  6.13 &  +2  2 54.8 & 0.138 & 18.15 & 17.01 & -20.45 &  -4.40 &  -6.49 &  -5.12 \\
290290 & TGN303Z261 & 11 13 27.92 &  -0 54 10.2 & 0.041 & 17.92 & 16.88 & -17.65 &  -5.57 &  -4.52 &  -5.88 \\
290537 & TGN312Z001 & 11 52 25.52 &  -1 16  3.3 & 0.061 & 17.97 & 16.85 & -18.55 &  -5.89 &  -5.42 &  -6.18 \\
294450 & TGN444Z229 & 11 33 49.63 &  +2  5 14.3 & 0.178 & 18.50 & 17.31 & -20.80 &  -6.46 &  -5.30 &  -5.77 \\
296025 & TGN449Z178 & 11 57 35.34 &  +2 10  3.7 & 0.003 & 16.46 & 15.86 & -13.52 &  -5.32 &  -6.44 &  -5.52 \\
321635 & TGN341Z170 & 14  9 43.20 &  -0  0 37.4 & 0.135 & 18.42 & 17.31 & -20.09 &  -5.12 &  -5.73 &  -5.35 \\
321865 & TGN274Z137 & 14  8 43.29 &  -1  9 41.3 & 0.005 & 14.10 & 13.06 & -16.93 &  -5.58 &  -6.29 &  -6.62 \\
326479 & TGN345Z098 & 14 30 55.85 &  +0  4 51.0 & 0.118 & 18.49 & 17.20 & -19.78 &  -4.62 &  -6.06 &  -6.45 \\
337194 & TGS867Z380 &  0 27  2.39 & -16 46 31.2 & 0.069 & 16.80 & 15.82 & -19.98 &  -6.18 &  -6.84 &  -6.92 \\
340630 & TGS842Z058 &  1  9 24.18 &  -7 37  5.1 & 0.092 & 19.17 & 17.90 & -18.42 &  -3.12 &  -9.57 &  -4.50 \\
350576 & TGS814Z259 &  2 18 38.37 & -45  4 44.7 & 0.071 & 16.22 & 15.07 & -20.68 &  -4.04 &  -4.98 &  -6.53 \\
360494 & TGS848Z072 &  4  0 56.87 & -35 57 14.1 & 0.167 & 18.78 & 17.62 & -20.34 &  -6.42 &  -5.85 &  -6.02 \\
386788 & TGS811Z409 & 23 32 35.56 & -41 40 49.7 & 0.227 & 19.06 & 17.56 & -21.14 &  -5.24 &  -6.37 &  -4.98 \\
\hline
\end{tabular}
\end{table*}

\end{document}